# Transport properties of spiral carbon nanofiber mats

# containing Pd metal clusters using Pd$_2$(dba)$_3$ as catalyst

C.-J. Liu[a],*, T.-W. Wu[a], L.-S. Hsu[a], C.-J. Su[b], C.-C. Wang[c], F.-S. Shieu[c]

[a]Department of Physics, National Changhua University of Education, Changhua 500, Taiwan, ROC

[b]Department of Applied Chemistry, Chung Shan Medical University, Taichung, Taiwan, ROC

[c]Department of Materials Engineering, National Chung Hsing University, Taichung, Taiwan, ROC

We have grown spiral carbon nanofibers containing Pd metal clusters using the Pd$_2$(dba)$_3$ catalyzed decomposition of gaseous acetylene on molecular sieves (AlPO$_4$-5) support.   The microstructure and composition of the spiral carbon nanofibers were examined by the powder x-ray diffractometer and transmission electron microscope.   The conductivity of the mat in the temperature range from 14 to 250 K could be described by the form of exp[-(T$^{-1/4}$)].   The thermopower shows a remarkably linear behavior down to 40 K, reminiscent of some conducting polymers.   The sign change of the thermopower suggests there exists more than one type of charge carrier, which could be ascribed to the different types of nanotube with various sizes of radius.   The transport behavior of spiral carbon nanofibers containing Pd metal clusters will be discussed in the framework of the heterogeneous model.

*Corresponding author: Tel: +886-4-723-2105; fax: +886-4-728-0698.  E-mail addrress: liucj@cc.ncue.edu.tw

# 1. Introduction

There have been extensive interests in the structure and morphology of carbon nanofibers prepared by the catalytic decomposition of gaseous hydrocarbon using various transition metals such as iron, cobalt, nickel, and some of their alloys [1-3].   These catalysts have been employed to engineer the conformation and crystalline perfection of carbon fibers and tubular structures of nano-sized dimensions [4].  These nanofibers were generally whisker-like tubular structures with diameters controlled by the size of the small metal particles.  In addition to the straight tubes, several intricate shapes and structures including helices, cones, tori, and rings have been identified [5-7].  At a higher structural level, strong inter-tube van der Waals (vdW) attraction can generate at least two kinds of self-assembled nanotubes: firstly, nanotubes can bunch together and lead to the rope formation [6]; secondly, upon ultrasonic irradiation, the nanotubes may be self-folded into ring configurations [8] and stabilized by the vdW attraction. However, very few observations of the self-organization among helically coiled nanotubes have been reported.

Based on the detailed plane-wave *ab initio* pseudopotentail local density function calculations, the electronic structure is sensitive to the diameter and chirality of the nanotubes, and hence the transport properties of the tubes could be drastically changed [9].  The small radius cabon nanotubes could obtain metallicity due to the $\pi^* - \sigma^*$ hybridization effects. Among the transport properties, the thermoelectric power (TEP) is known to be a sensitive probe of the energy dependence of the charge carrier relaxation processes and topology of the Fermi surface.   TEP measurements have been used to characterize both the single-walled carbon nanotubes and the multi-walled carbon nanotubules bundles which contain no Pd clusters and almost certainly have very few spiral nanotues [10,11].  We have recently prepared new material of spiral carbon nanofibers containing Pd metal clusters, it would therefore be interesting to characterize this new material using TEP measurements. In this work, we report the preparation, TEM observations, electrical resistivity, and TEP measurements of the spiral carbon nanofibers containing Pd metal clusters. We find that the temperature dependence of the conductivity between 14 and 250 K could be described by the variable-range-hopping process, namely, the Mott's $T^{-1/4}$-law. The thermopower exhibits a remarkably linear behavior down to 40 K, reminiscent of some conducting polymers. Together with the observation of the microstructure of the sample, the transport behavior seems to indicate the sample containing spiral nanofibers having different pitches and radius can be explained in the scenario of the heterogeneous model [12].

## 2. Experiment

The synthesis of the spiral carbon nanofibers containing Pd metal clusters is briefly described as follows. The preparation of the sample is described as follows. The carbon nanofibers were synthesized using the $AlPO_4$-5 (Strem) supported $Pd_2(dba)_3$ (Aldrich) catalyst, which was prepared by the impregnation method. $Pd_2(dba)_3$ fine powder of 0.1g was dissolved in 10 ml distilled water and stirred for 10 min. at 70°C. One gram of microporous $AlPO_4$-5 was then slowly added into the solution and stirred for 1 h. The mixture was ground to fine purple-gray powder (100 mesh) and drying in an oven at 115°C for 10 h. The resulting catalyst was heated to 200°C in air and then rinsed at 200°C for 3 h. The gas mixture ($N_2$ = 100 ml min$^{-1}$ and $C_2H_2$ = 10 ml min$^{-1}$) was introduced for about 10 min before the sample was brought to 700°C, and the reaction was allowed to proceed at 700°C for another 1.5 h, followed by furnace cooling for 15 h.  The reaction product was then collected as fine black powders.  It is noted that the synthesis of spiral carbon nanofibers is very sensitive to the reaction temperature. Only within a narrow temperature range of 700°C have the spiral structures been observed by the current method of preparation. No spiral nanofiber was observed at 650°C; and at 750°C, very few spiral nanotubes were observed by TEM.

The powder x-ray diffraction (XRD) patterns were obtained by using a powder X-ray diffractometer equipped with the Cu $K_\alpha$ radiation.  The microstructure of the carbon nanofibers was observed using TEM (Hitachi H-7100, 120 kV) after sonicating the samples in ethanol for 1 h and dispersing a drop of solution over holey carbon grids.  The as-prepared powders were cold-pressed with pressure about $8.5 \times 10^4$ N/cm$^2$ and made into parallelepiped pellets without further heat treatments, which results in a density of 0.11g/cm$^3$.  The temperature dependence of dc electrical resistivity was measured using standard four-probe techniques in the temperature range of 10 - 300 K.  Silver paste was applied to make the electrical contacts.  The magnetoresistance measurements were carried out with the applied field parallel to the applied current.  TEP was measured between 10 and 300 K using an Oxford closed cycle cooler cryostat.  A Cernox sensor was used to monitor the ambient temperature of the sample.  TEP data were only collected after the system reached a steady state, which was monitored by a user-written program.  The temperature gradient across the sample was monitored using two type E thermocouples connected in a differential mode.   Temperature gradients were typically between 0.5 and 1 K.  The TEP of the sample was obtained by subtracting the TEP of Cu Seebeck probes.

## 3. Results

As shown in Fig. 1, the XRD pattern shows the peak broadening of carbon at $2\theta = 26°$ and $44°$, clearly indicating the small particle size effect of the carbon nanofibers.  Several low-indexed Pd peaks also show peak broadening, suggesting the presence of Pd cluster in the nanofibers.  Besides the irregularly curved carbon nanotubes, the TEM image shown in Fig. 2 clearly displays three major types of spiral structure of the carbon nanofibers, which are distinguished according to their pitch (p) to diameter (2r) ratio, or equivalently, the pitch angle of the helix, $\varphi = \arctan(p / 2\pi r)$ [13].  The pitch for each type of nanofiber is not easy to quantify, nevertheless, the diameter of the three types of nanofibers are estimated to be 135 nm, 80 nm and 43 nm, respectively.  Since the low pitch-angle nanofibers appear to have larger vdW adhesive energy due to large contact area among different strands, it is proposed here that the vdW adhesive energy is the major factor for the stability of the spiral structure of the carbon nanofibers.  Fig. 3 shows the result of energy-dispersive x-ray (EDX) analyses on the rectangular areas indicated in Fig. 3 and provides evidence that the black spots inside the carbon nanofibers are palladium metal clusters.  This observation strongly supports the successful incorporation of Pd metal cluster inside the carbon nanofibers in our synthetic processes.

Fig. 4 shows the temperature dependence of resistivity of the mat down to 10 K in zero field and H = 5000 Oe for our sample.  Conduction will clearly take place along the connected spiral nanotubes fibers as shown in Fig. 2.  Although the magnitude of resistance is greatly affected by the number of fibers per unit area of cross-section in the sample, the temperature and magnetic field dependence are characteristic of the conduction process along the connected fibers and indicate the conduction mechanisms [14].  Since the density of 0.11g/cm$^3$ in our sample is considerably less than the density 2.26g/cm$^3$ of graphite, it confirms that we have a porous network of spiral carbon fibers.  Note that the plotted resistivity in Fig. 4 is that of the sample as a whole and not the magnitude of the intrinsic resistivity of individual fibers.  It is clear that the temperature dependence of resistivity is of nonmetallic type over the whole investigated temperature range.  The high resistivity and its relatively small increase with decreasing temperature are indicative of a disordered conductor.  In Fig. 5, we plot ln$\sigma$ versus $T^{-1/4}$ on a semilog scale between 10 and 250 K from the resistivity data given in Fig. 4.  The good fit of ln$\sigma$ to $T^{-1/4}$ in the range of 14 – 250 K might suggest a variable-range hopping (VRH) process dominating the transport.  The fits to the $T^{-1/3}$ and $T^{-1/2}$ laws are not as linear as the fits to $T^{-1/4}$.  However, a deviation from Mott's $T^{-1/4}$- law tends to develop below ~14 K.  Fig. 6 shows the field dependence of the MR ratio, defined as [ $\rho$ (H)- $\rho$ (0)]/ $\rho$ (0), up to 5000 Oe at various temperatures.  The MR ratio lies in the range of –3 to +3 % for H = 1000 - 5000 Oe.  At H = 1000 Oe, the MR ratio is positive, whereas for H > 1000 Oe the MR ratio is negative (except at T = 300K).  Between T = 30 and 190 K, the magnitude of MR increases gradually with increasing temperature (not shown).    It is clearly seen that the MR at all temperatures shows an upturn at 2000 Oe.  For a localized system with variable-range hopping transport, destruction of the phase coherence between different hopping paths in a magnetic field could lead to negative magnetoresistance [15].

Fig. 7 shows the temperature dependence of TEP in zero field for the carbon nanofibers containing Pd metal cluster.  Thermopower is an intrinsic property of materials that is not greatly affected by sample directions, alignment of fibers or even by thin barriers in conducting fibers [14].  The features of our TEP data are described as follows.  The temperature dependence of TEP in zero field shows a remarkably linear behavior for T > 40 K followed by a rapid increase in absolute value for T < 40 K.  The TEP also exhibits a sign crossover at 225 K, suggesting there is more than one type of charge carrier existing in the system.  Note that the room-temperature thermopower of single-walled carbon nanotubes network (without Pd clusters or spiral nanotubes) was found to be quite large (in the range of 20-60 $\mu$V/K) as compared to our sample and remained positive down to 4.2 K [10,11,16].  The smaller absolute value of measured TEP and sign crossover in our sample suggest compensation of electrons and holes involved in transport.

## 4. Discussion

According to the band structure calculations, it has been shown that the electronic property of graphite nanotubules could be a good metal or a semiconductor with various energy gaps due to the variation of diameter and the degree of helical arrangement in the geometrical structure [17-19].  For semiconducting microtubules with larger radius, a larger energy gap is predicted.  From Fig. 2, the TEM image shows that there are at least 3 types of spiral structure having different diameters and pitches in our carbon nanofibers containing Pd clusters.  The arrangement of different types of spiral structures in the sample is somewhat similar to a random resistor network system.   The microstructure could be considered as a cluster composed of different types of spiral nanofibers.  Apparently, the conduction path is not a single channel with one type of connection between nanofibers.  Instead, it is more of a network of resistors.  Some of the nanofibers are linked to each other with either one connection or multiple connections.  Some of them have one end linked to other nanofibers and no connection at the other end, resembling a dangling "bond" where the current path is blocked [20].

In variable-range-hopping transport, the conductivity would follow the Mott's law,

$$\sigma = \sigma_0 \exp\left[-\left(\frac{T_0}{T}\right)^v\right],$$

(1)

where $\sigma_0$ is weakly temperature dependent but barely affects the value of the index $v$, and $T_0$ is associated with localization length.  For a 3-dimensional system, the index $v$ is 1/4, provided the density of states near $E_F$ is constant.

The nonmetallic temperature dependence of electrical resistivity could arise from the fact that three types of nanofibers coexist but possess different resistivities.  In describing the conduction process of such a network of resistors, one might make use of the heterogeneous model.  In a heterogeneous model, the total electrical resistivity $\rho_T$ is given by [21]

$$\rho_T = \sum_i R_i \frac{A_i}{L_i},$$

(2)

where $L_i$ is the fraction of sample length for the materials with electrical resistance $R_i$, and $A_i$ the corresponding fraction of sample cross-section area.  The total electrical resistivity is then dominated by the more resistive inter-fibril contact barriers, which might account for the nonmetallic temperature-dependent behavior in our sample.    Furthermore, the good fit of Mott's $T^{-1/4}$-law seems to support the view that our sample resistance is dominated by variable-range hopping.

We now turn to the TEP data given in Fig. 7. The temperature dependence of thermopower with temperatures above 40 K for our material is close to metallic diffusion thermopower. This situation seems not to reconcile with the nonmetallic behavior in the temperature dependence of electrical resistivity. In fact, this behavior is reminiscent of the organic conducting polymers (polyaniline and plypyrrole) and polyaniline blends [22-24]. The metallic diffusion thermopower behavior may result from the fact that the inter-fibril contact barriers have little contribution to thermopower but do have effects on the electrical resistivity. According to Kohler's formula, the diffusion thermopower $S_D$ for a conductor made of $i$ elements in series can be expressed as

$$S_D = \frac{\sum_i S_i W_i}{\sum_i W_i}, \tag{3}$$

where $S_i$ is the partial thermopower of the $i^{th}$ element and $W_i$ the thermal resistance. The inter-fibril contact barriers make our material show nonmetallic behavior in the temperature dependence of the electrical resistivity, but the small weighting factor of thermal resistivity due to the inter-fibril contact barriers would not affect the metallic diffusion thermopower behavior. Furthermore, a sign crossover of thermopower occurs at $T = 225$ K. This result suggests that there exists more than one type of carrier in our sample. The sign crossover may arise from the three types of carbon nanofibers containing Pd metal cluster observed in the TEM image having opposite sign of carriers. The magnitude of thermopower increases dramatically from – 2.53 $\mu$ V/K at 40 K to −12.9 $\mu$ V/K at 15 K.   The characteristic thermopower for a semiconductor with an energy gap is increasing in absolute value with decreasing temperature. The slope change in the temperature dependence of thermopower indicates the dominating transport at T < 40 K switches to a different mechanism.  This region corresponds to strongly increasing resistivity.

For a disordered system with variable-range-hopping transport between localized states, the TEP should follow the form [25]

$$S(T) \propto T^{\frac{d-1}{d+1}},  \tag{4}$$

where $d$ is the dimensionality of the system.  The TEP should follow the form of of $T^{1/2}$ for 3D, and of $T^{1/3}$ for 2D.  As seen in Eq. (4), the TEP would decrease with decreasing temperature and tends to zero as the temperature approaches zero because of the energy collapsing of carriers.  However, this is not the case shown in Fig. 7.  When compared with the experimental data in general, the $T^{1/2}$ law is not well supported although the TEP of nonsuperconducting $La_{1.8}Sr_{0.2}CaCu_2O_{6-\delta}$ without hot-isostatic-press annealing [26] is nicely fitted to $T^{1/2}$ law and some of the conducting polymers could fit to a linear T plus $T^{1/2}$ term ($S = XT + CT^{1/2}$) [27]. The linear behavior of TEP for T > 40 K could be attributed to the hole carriers of metallic nanotubes; the diverging behavior of TEP below 40 K suggests a change to dominant semiconductor behavior.

## 5. Conclusions

We have measured the transport properties of spiral carbon nanofibers containing Pd metal clusters grown by the $Pd_2(dba)_3$-catalyzed decomposition of gaseous acetylene on molecular sieves ($AlPO_4$-5) support.  The XRD pattern and the TEM image show that the sample is carbon nanofibers containing Pd metal clusters.  In combination with the observation of the microstructure of the sample, the transport behavior seems to indicate the sample consisting of spiral nanofibers having different pitches and radius can be explained in the scenario of the heterogeneous model.

Acknowledgments

  This work was supported by the National Science Council, Taiwan, ROC under the grants of NSC 91-2112-M-018-011 and NSC-91-2112-M-018-007.

References

 [1] Dresselhaus MS, Dresselhaus G, Eklund PC. Science and technology of fullerenes and carbon Nanotubes. San Diego: Academic Press. 1996.

[2] Rodriguez NM. A review of catalytically grown carbon nanofibers. J Mater Res 1993; 8(12):3233-3250.

[3] Endo M, Takeuchi K, Igarashi S, Kobori K, Shiraishi M, Kroto HW. The production and structure of pyrolytic carbon nanotubes. J Phys Chem Solids 1993; 54(12):1841-1848.

[4] Rodriguez NM, Chambers A, Baker RTK. Catalytic engineering of carbon nanostructures. Langmuir 1995; 11(10):3862-3866.

[5] Amelinckx S, Zhang XB, Bernaerts D, Zhang XF, Ivanov V, Nagy JB. A formation mechanism for catalytically grown helix-shaped graphite nanotubes. Science 1994; 265: 635-639.

[6] Thess A, Lee R, Nikolaev P, Dai H, Petit P, Robert J, et al. Crystalline Ropes of Metallic Carbon Nanotubes. Science 1996; 273:483-7.

[7] Gao R, Wang Z-L, Fan S. Kinetically controlled growth of helical and zigzag shapes of carbon nanotubes. J Phys Chem B 2000; 104 (6):1227-1234.

[8] Martel R, Shea HR, Avouris P. Ring Formation in Single-Wall Carbon Nanotubes. J Phys Chem B 1999; 103(36):7551-7556.

[9] Blasé X, Benedict LX, Shirley EL, Louie SG. Hybridization effects and metallicity in small radius carbon nanotubes. Phys Rev Lett 1994; 72(12):1878-1881.

[10] Hone J, Ellwood I, Muno M, Mizel A, Cohen ML, Zettl A., et al. Thermoelectric power of single-walled carbon nanotubes. Phys Rev Lett 1998; 80(5):1042-5.

[11] Tian M, Li F, Chen L, Mao Z, Zhang Y. Thermoelectric power behavior in carbon nanotubule bundles from 4.2 to 300 K. Phys Rev B 1998; 58(3):1166-8.

[12] Kaiser AB, Düsberg G, Roth S. Heterogeneous model for conduction in carbon nanotubes. Phys Rev B 57 1998; 57(3):1418-1421.

[13] Zhong-can O-Y, Su ZB, Wang CL.  Coil formation in multishell carbon nanotubes: competition between curvature elasticity and interlayer adhesion. Phys Rev Lett 1997; 78(21): 4055-4058.

[14] Kaiser AB, Electronic transport properties of conducting polymers and carbon nanotubes. Rep. Prog. Phys. 2001, 64: 1-49.

[15] Sivan U, Entin-Wohlman O, Imary Y. Orbital magnetoconductance in the variable-range–hopping regime. Phys Rev Lett 1988; 60(15):1566-9.

[16] Grigorian L, Williams KA, Fang S, Sumanasekera GU, Loper AL, Dickey EC, et. al. Reversible intercalation of charged iodine chains into carbon nanotube ropes. Phys Rev Let 1998; 80(25):5560-3.

[17] Hamada N, Sawada S-I, Oshiyama A. New one-dimensional conductors: Graphitic microtubules. Phys Rev Lett 1992; 68(10):1579-1581.

[18] Yorikawa H, Muramatsu S. Electronic properties of semiconducting graphitic microtubules. Phys Rev B 1994; 50(16):12203-6.

[19] White CT, Robertson DH, Mintmire JW. Helical and rotational symmetries of nanoscale graphitic tubules.  Phys Rev B 1993; 47(9):5485-8.

[20] Stauffer D, Aharony A. Introduction to percolation theory. London: Taylor & Francis. 1994.

[21] Kaiser AB, Flanngan GU, Stewart DM, Beaglehole D. Heterogeneous model for conduction in conducting polymers and carbon nanotubes. Synth Met 2001; 117(1-3): 67-73.

[22] Subramaniam CK, Kaiser AB, Gilberd PW, Liu CJ, Wessling B. Conductivity and thermopower of blends of polyaniline with insulating polymers (PETG and PMMA). Solid State Commun 1996; 97(3):235-8.

[23] Kaiser AB, Liu CJ, Gilberd PW, Chapman B, Kemp NT, Wessling B, et al. Comparison of electronic transport in polyaniline blends, polyaniline and polypyrrole. Synth Met 1997; 84(1-3):699-702.

[24] Kemp NT, Kaiser AB, Liu CJ, Chapman B, Mercier O, Carr AM, et al. Thermoelectric power and conductivity of different types of polypyrrole. J Polym Sci Part B: Polym Phys 1999; 37(9):953-960.

[25] Mott NF, Davis EA. Electronic processes in noncrystalline materials, $2^{nd}$ ed. Oxford: Oxford University Press. 1979:55.

[26] Liu CJ, Yamauchi H. Thermoelectric power and resistivity of $La_{1.8}Sr_{0.2}CaCu_2O_{6-\delta}$ and the effects of $O_2$ hot-isostatic-press annealing. Phys. Rev. B 1995; 51(17):11826-9.

[27] Kaiser AB. Thermoelectric power and conductivity of heterogeneous conducting polymers. Phys Rev B 1989; 40(5):2806-2813.

Figures Captions

FIG. 1. The X-ray diffraction pattern of carbon nanofibers containing Pd metal clusters.

FIG. 2. A TEM image showing three types of carbon nanofibers containing Pd metal clusters, labeled by A, B, and C, respectively.  The square indicates the presence of the Pd metal cluster, evidenced by the energy-dispersive x-ray analyses.

FIG.3. The energy-dispersive x-ray analysis of the rectangular areas indicated in Fig. 2, showing the presence of Pd metal clusters in spiral carbon nanofibers.  The peak of Cu is due to the Cu grids used in the analysis.

FIG. 4. The temperature dependence of electrical resistivity in zero field and H = 5000 G for carbon nanofibers containing Pd metal clusters.

FIG. 5. The conductivity $\ln(\sigma)$ at $9.8 \leq T \leq 250.2$ K replotted against $T^{-1/4}$ for carbon nanofibers containing Pd metal clusters; the good fit of $\ln\sigma$ to $T^{-1/4}$ in the range of 14 – 250 K suggest a variable-range hopping process for carriers percolating in a scattering medium consisting of randomly distributed scattering centers.  R is the correlation coefficient.

FIG. 6. The field dependence of MR ratio for the carbon nanofibers containing Pd metal clusters.  At H = 1000 Oe, the MR ratio is positive, whereas for H > 1000 Oe the MR ratio is negative (except at T = 300K).  The solid line is a guide for the eyes.

FIG. 7.  The temperature dependence of the TEP for the carbon nanofibers containing Pd metal clusters shows a remarkably linear behavior down to 40 K, reminiscent of some organic conducting polymers and blends.

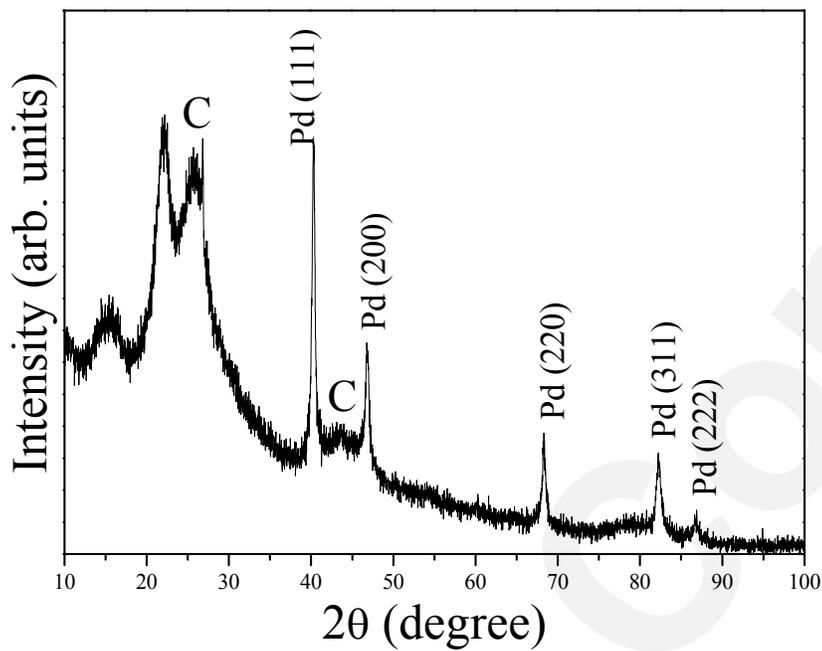

FIG. 1. The X-ray diffraction pattern of carbon nanofibers containing Pd metal clusters.

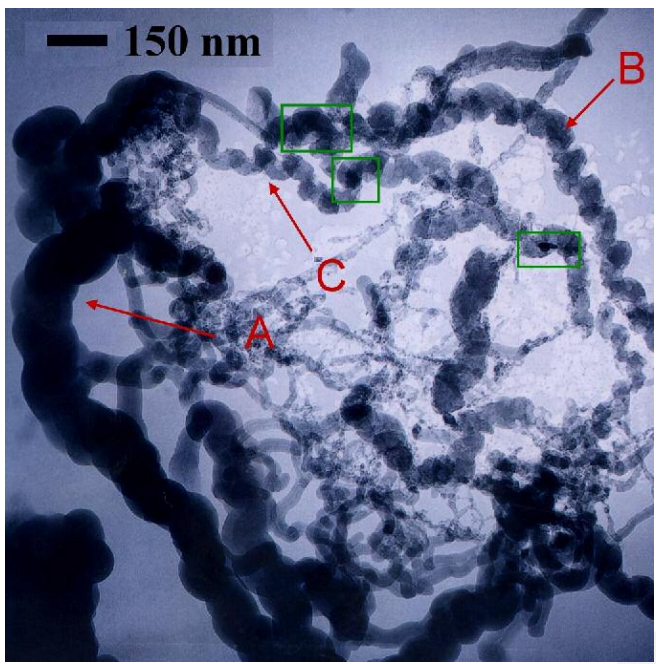

FIG. 2. A TEM image showing three types of carbon nanofibers containing Pd metal clusters, labeled by A, B, and C, respectively. The square indicates the presence of the Pd metal cluster, evidenced by the energy-dispersive x-ray analyses.

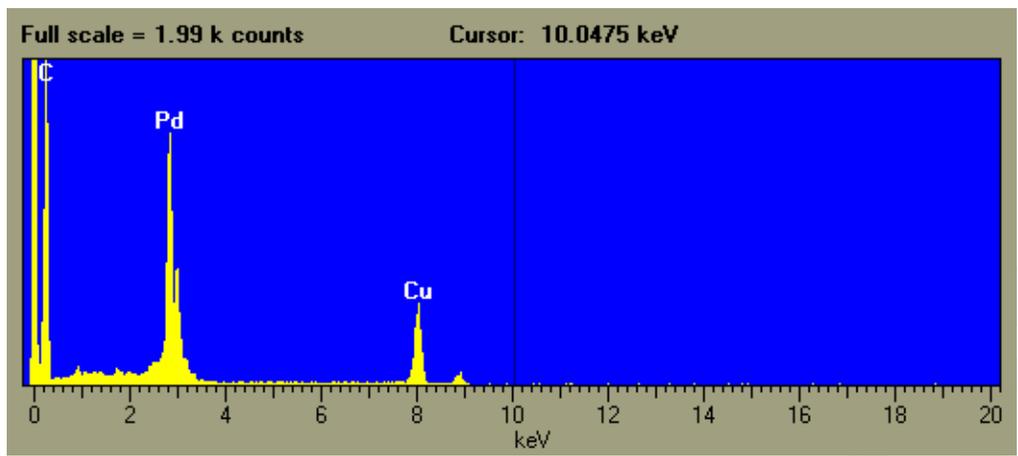

FIG.3. The energy-dispersive x-ray analysis of the rectangular areas indicated in Fig. 2,

showing the presence of  Pd metal clusters in spiral carbon nanofibers.  The peak of

Cu is due to the Cu grids used in the analysis.

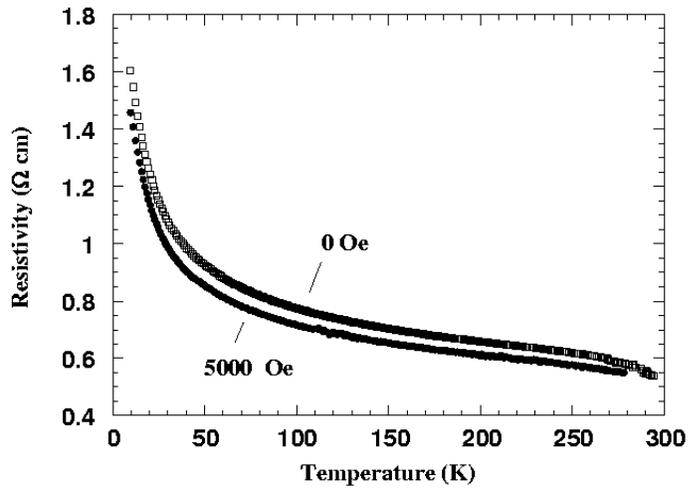

FIG. 4. The temperature dependence of electrical resistivity in zero field and H = 5000 G

for carbon nanofibers containing Pd metal clusters.

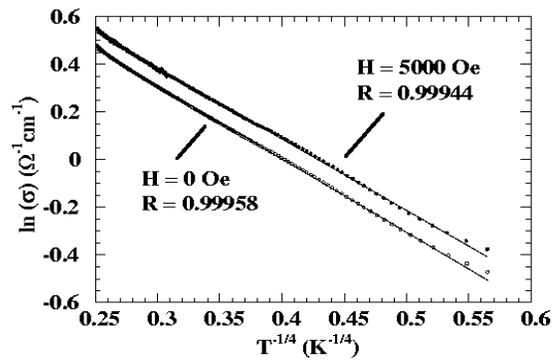

FIG. 5. The conductivity ln(σ) at $9.8 \leq T \leq 250.2$ K replotted against $T^{-1/4}$ for carbon nanofibers containing Pd metal clusters; the good fit of lnσ to $T^{-1/4}$ in the range of $14 - 250$ K suggest a variable-range hopping process for carriers percolating in a scattering medium consisting of randomly distributed scattering centers.  R is the correlation coefficient.

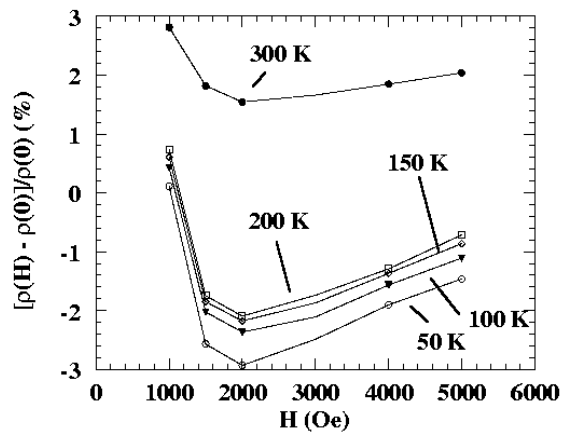

FIG. 6. The field dependence of MR ratio for the carbon nanofibers containing Pd metal

clusters.  At H = 1000 Oe, the MR ratio is positive, whereas for H > 1000 Oe the

MR ratio is negative (except at T = 300K).  The solid line is a guide for the eyes.

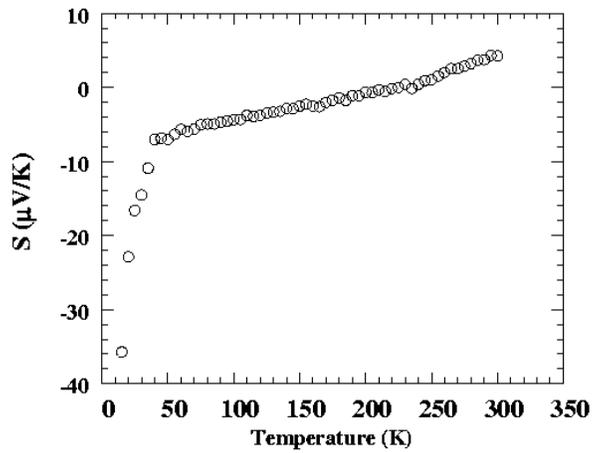

FIG. 7.  The temperature dependence of the TEP for the carbon nanofibers containing Pd

metal clusters shows a remarkably linear behavior down to 40 K, reminiscent of some organic conducting polymers and blends.